\documentclass[twocolumn,showpacs,preprintnumbers,amsmath,amssymb,aps,pre,floatfix,superscriptaddress]{revtex4-1}

\usepackage{graphicx}
\usepackage{dcolumn}
\usepackage{bm}
\usepackage[utf8]{inputenc}
\usepackage{booktabs}
\usepackage[Gray,thinqspace,squaren]{SIunits}
\usepackage{textcomp}

\begin{document}

\newcommand{\bs}[1]{{\boldsymbol #1}}
\newcommand{\mb}[1]{{\mathbf #1}}
\renewcommand{\l}{\left}
\renewcommand{\r}{\right}
\newcommand{\mi}{\mathrm i}
\newcommand{\me}{\mathrm e}
\newcommand{\md}{\mathrm d}
\newcommand{\KM}[1]{D^{(#1)}}
\newcommand{\6}[2]{\frac{\partial #1}{\partial #2}}

\title{Extended Kramers-Moyal analysis applied to optical trapping}

\author{Christoph Honisch}
\email{c.honisch@uni-muenster.de}
\author{Rudolf Friedrich}
\affiliation{Institute for Theoretical Physics, University of Muenster, D-48149 Muenster, Germany}

\author{Florian H\"orner}
\affiliation{Institute of Cell Biology, ZMBE, D-48149 Muenster, Germany}
\affiliation{Institute for Applied Physics, University of Muenster, D-48149 Muenster, Germany}

\author{Cornelia Denz}
\affiliation{Institute for Applied Physics, University of Muenster, D-48149 Muenster, Germany}

\date{\today}

\begin{abstract}
The Kramers-Moyal analysis is a well established approach to analyze stochastic time series from complex systems. If the sampling interval of a measured time series is too low, systematic errors occur in the analysis 
results. These errors are labeled as finite time effects in the literature. In the present article, we present some new insights about these effects and discuss the limitations of a previously published method to estimate Kramers-Moyal coefficients at the presence of finite time effects. To increase the reliability of this method and to avoid misinterpretations, we extend it by the computation of error estimates for estimated parameters using a Monte Carlo error propagation technique. Finally, the extended method is applied to a data set of an optical trapping experiment yielding estimations of the forces acting on a Brownian particle trapped by optical tweezers. We find an increased Markov-Einstein time scale of the order of the relaxation time of the process which can be traced back to memory effects caused by the interaction of the particle and the fluid. 
Above the Markov-Einstein time scale, the process can be very well described by the classical overdamped Markov model for Brownian motion.

\end{abstract}


\maketitle

\section{Introduction}

Many real world stochastic processes $q(t)$ can be modeled by stochastic differential equations of Langevin type  \cite{risken,franck2010book,vkampen2007book},
\begin{equation} \label{eq:langevin}
  \dot q = \KM 1(q) + \sqrt{2\KM 2(q)}\varGamma\,,
\end{equation}
where $\KM 1(q)$ is called the drift coefficient that describes deterministic influences on the dynamics and $\KM 2(q)$ is the diffusion coefficient that describes the state dependent amplitude of a fast fluctuating force $\varGamma(t)$ with zero mean, adopting It\^{o}'s interpretation. A frequently applied idealization is to assume that $\varGamma(t)$ is $\delta$ correlated in time and Gaussian distributed. In this case the process fulfills the Markov property and can be completely characterized by the corresponding Fokker-Planck equation (FPE)
\begin{equation} \label{eq:FPE}
  \6{}{t}f_q(x,t) = \hat L(x) f_q(x,t)
\end{equation}
with the Fokker-Planck operator
\begin{equation}
  \hat L(x)=\l[-\6{}{x}\KM 1(x) + \6{^2}{x^2}\KM 2(x)\r]
\end{equation}
that describes the temporal evolution of the probability density function (PDF) $f_q(x,t)=\langle \delta(x-q(t)) \rangle$ of the process $q(t)$. The Fokker-Planck operator $\hat L(x)$ depends on the drift and diffusion coefficients that determine the corresponding Langevin equation \eqref{eq:langevin}.

Drift and diffusion coefficients are also referred to as the first and second Kramers-Moyal (KM) coefficients, respectively. Given the transition probability densities $p_q(x',t+\tau|x,t)$ of a general Markov process $q(t)$, the $n$th KM coefficient can be defined as
\begin{equation}
  \KM n(x) = \lim_{\tau \rightarrow 0} \frac{1}{n!\tau} M^{(n)}_{\tau}(x) \,,
\end{equation}
where $M^{(n)}_{\tau}(x)$ is called the $n$th conditional moment given by
\begin{equation} \label{eq:cond_mom}
\begin{split}
  M^{(n)}_{\tau}(x) &=\langle (q(t+\tau)-q(t))^n \rangle|_{q(t)=x} \\
                    &= \int (x'-x)^n p_q(x',t+\tau|x,t) \md x'\,.
\end{split}
\end{equation}
Since the transition PDFs can be estimated from measured time series data of a process of interest, it is in principle possible to set up a model in terms of a Fokker-Planck or Langevin equation by data analysis. Since Friedrich and Peinke applied this nowadays called KM analysis to the investigation of the turbulent cascade \cite{friedrich1997pd,friedrich1997prl} in 1997, it has developed to a rapidly growing field of research with many applications in the natural sciences and beyond (see \cite{friedrich2011pr} and references therein).

One major problem connected to the KM analysis is the limit of the time increment $\tau$ to zero that has to be performed in the determination of the KM coefficients. Without an appropriate limiting procedure, estimated KM coefficients can deviate significantly from the true coefficients if the minimal available $\tau$, i.\,e. the sampling interval of the measured time series, is too large. These errors are referred to as finite time effects and were the subject of many publications during the last years (e.\,g. \cite{ragwitz2001prl,friedrich2002prl,lade2009pla,anteneodo2010pre,riera2010jsm}). Without an adequate understanding of finite time effects, KM analysis involves the risk of dramatic misinterpretations of the achieved results.

In a recent publication \cite{honisch2011pre}, we have developed a method that allows for a correct KM analysis when the sampling interval is large. In the present article we demonstrate that also this method fails if the sampling interval is so large that the information about the true KM coefficients is (almost) lost. In order to lower the risk of misinterpretations in those cases, we extend the method by the computation of error estimates for the obtained model parameters. 
The problem of error estimates for model parameters in the KM analysis context is also addressed in a very recent publication by Kleinhans \cite{kleinhans2012pre} in a Bayesian framework. In contrast to our approach, this method is only designed for data sets with a sufficiently large sampling rate.

We present an application of our method to an optical experiment that yields real-world stochastic data via trajectories of a Brownian particle trapped by optical tweezers.
We apply our method to these trajectories in order to yield estimations for the forces induced by the optical tweezers and to analyze the spatial distribution of the temperature.

The outline of the article is as follows. In Sec. \ref{sec:fte} we review the problem of finite time effects and discuss an example in order to provide an intuitive understanding. Afterwards we review the method of Ref. \cite{honisch2011pre} and present important extensions. Finally, Sec. \ref{sec:application} describes the application to our experiment. The last section is dedicated to some concluding remarks.

\section{Finite time effects} \label{sec:fte}

At first we introduce the finite time KM coefficients
\begin{equation} \label{eq:def_ftc}
  \KM n_{\tau}(x) = \frac{1}{n!\tau} M^{(n)}_{\tau}(x)\,,
\end{equation}
where $M^{(n)}_{\tau}(x)$ are the conditional moments defined in Eq. \eqref{eq:cond_mom}. Apparently, $\lim_{\tau \rightarrow 0} \KM n_{\tau}(x) = \KM n(x)$. In order to understand finite time effects and to possibly correct for them, it is important to know, how the finite time coefficients change with the time increment $\tau$ given the true coefficients. 

There are several possibilities to compute finite time coefficients. One possibility is by solving the FPE. The transition PDF $p_q(x,t_0+\tau|x_0,t_0)$ that occurs in Eq. \eqref{eq:cond_mom} is the solution to the FPE \eqref{eq:FPE} at time $t=t_0+\tau$ with the initial condition 
\begin{equation}
  f_q(x,t_0) = \delta(x_0)\,.
\end{equation}
Unfortunately this is in most cases impossible to do analytically. Also for a numerical solution of the FPE, the initial condition in form of a Dirac $\delta$ distribution will cause problems.

Another possibility is to numerically integrate the corresponding Langevin equation and to estimate the finite time coefficients from the generated time series. Based on this approach, Kleinhans \emph{et al.} developed a maximum likelihood approach to estimate KM coefficients for processes with finite sampling rates \cite{kleinhans2005pla,kleinhans2007pla}.

A third possibility is the adjoint operator approach. In Ref. \cite{friedrich2002prl}, Friedrich \emph{et al.} mention that the conditional moments can be expressed as a Taylor series in $\tau$ which reads
\begin{equation} 
  M^{(n)}_{\tau}(x) = \l.\l[\sum_{k=0}^{\infty} \frac{ \l(\hat L^{\dag}(x')\r)^k \tau^k}{k!} \l(x'-x\r)^n \r] \r|_{x'=x}\,. \label{eq:m_taylor}
\end{equation}
This series expansion contains the adjoint Fokker-Planck operator
\begin{equation}
  \hat L^{\dag}(x') = \KM 1(x') \6 {}{x'} + \KM 2(x') \6{^2}{x'^2}\,.
\end{equation}
With the adjoint operator method it is not only possible to obtain a series expansion. As Lade showed in Ref. \cite{lade2009pla}, the conditional moments can also be computed by solving the partial differential equation
\begin{equation}
  \6{W_{n,x}(x',t)}{t} = \hat L^{\dag}(x') W_{n,x}(x',t)\,, \label{eq:AFPE}
\end{equation}
with initial condition
\begin{equation}
  W_{n,x}\l(x',0\r) = \l(x' - x\r)^n\,. \label{eq:AFPE_IC}
\end{equation}
Then the conditional moments are given by
\begin{equation}
  M^{(n)}_{\tau}(x) = W_{n,x}\l(x'=x,t=\tau\r)\,.
\end{equation}
Eq. \eqref{eq:AFPE}, also called the adjoint FPE (AFPE), can easily be solved analytically as long as the drift is linear. Otherwise numerical solutions can be obtained with standard finite difference schemes.

For an Ornstein-Uhlenbeck (OU) process,
\begin{subequations} \label{eq:OU}
\begin{align}
  \KM1(x) &= -\gamma x\,, \\
  \KM2(x) &= \alpha \,,
\end{align} 
\end{subequations}
the analytic expressions for the finite time coefficients read
\begin{subequations} \label{eq:OU_FT}
\begin{align} 
  \KM 1_{\tau}(x) &= -\frac{x}{\tau} \l(1-\me^{-\gamma \tau}\r)\,, \label{eq:drift_OU}\\
  \KM 2_{\tau}(x) &= \frac 1{2\tau}\l[ x^2\l(1-\me^{-\gamma \tau} \r)^2 + \frac \alpha{\gamma} \l( 1-\me^{-2\gamma \tau} \r) \r]\,.
\end{align}
\end{subequations}

\begin{figure}[htb]
  \includegraphics[width = 0.45\textwidth]{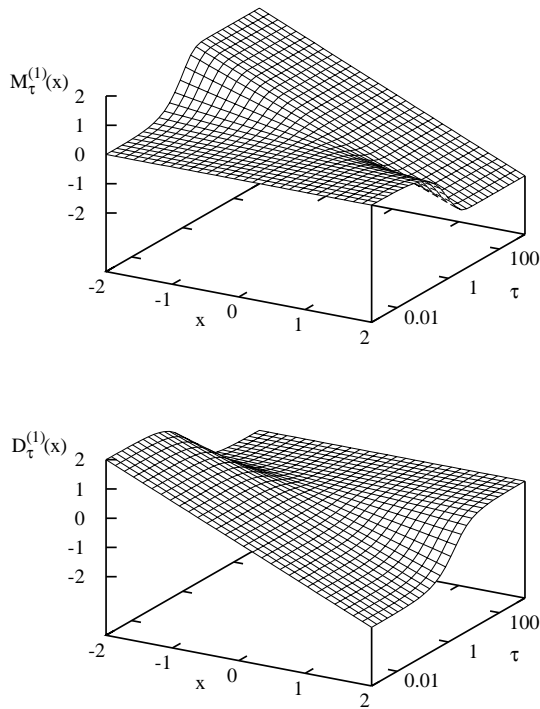}
  \caption{Conditional moment (top) and finite time drift (bottom) for an OU process with $\KM 1=-x$ and $\KM 2=1$. The $\tau$ axes are in logarithmic scaling.} \label{fig:OU_exmpl_1}
\end{figure}

\begin{figure}[htb]
  \includegraphics[width = 0.45\textwidth]{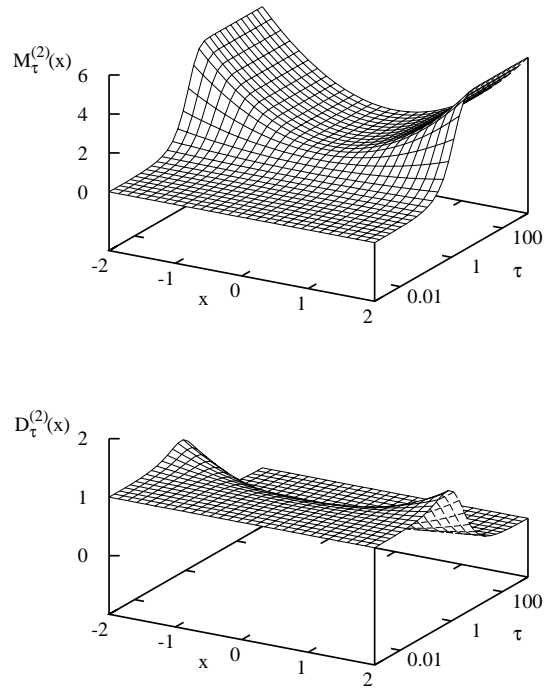}
  \caption{Conditional moment (top) and finite time diffusion (bottom) for the same process as in Fig. \ref{fig:OU_exmpl_1}.} \label{fig:OU_exmpl_2}
\end{figure}

We have plotted the finite time coefficients together with the conditional moments in Fig. \ref{fig:OU_exmpl_1} (drift) and Fig. \ref{fig:OU_exmpl_2} (diffusion) for $\gamma=\alpha=1$. By taking a logarithmic scaling for the $\tau$ axis, one can clearly identify two limiting cases separated by the relaxation time, which is $\tau_R=1$ in this case.
For $\tau\ll\tau_R$, the finite time KM coefficients have converged to the true KM coefficients, whereas the conditional moments tend to zero. For $\tau\gg\tau_R$, the conditional moments become stationary with respect to $\tau$, whereas the corresponding KM coefficients vanish.

The latter case is the limit of statistical independence that has been discussed by Anteneodo and Queir\'{o}s in Ref. \cite{anteneodo2010pre}. For $\tau\gg\tau_R$, the realization $x$ of a process $q$ at time $t$ becomes statistically independent from the realization $x'$ at time $t+\tau$. Therefore,
\begin{equation}
  p_q(x',t+\tau|x,t) \approx f_q(x',t+\tau)\,.
\end{equation}
For a stationary process, $f_q(x',t+\tau)$ is just the stationary PDF $f_q(x)$. As a consequence, the conditional moments yield
\begin{subequations} \label{eq:m_si}
\begin{align}
  M^{(1)}_{\tau\gg\tau_R}(x) &\simeq \langle q \rangle - x\,, \\
  M^{(2)}_{\tau\gg\tau_R}(x) &\simeq \langle q^2 \rangle - 2\langle q\rangle x + x^2\,.
\end{align}
\end{subequations}
and the finite time KM coefficients read
\begin{subequations} \label{eq:D12si}
\begin{align}
  D^{(1)}_{\tau\gg\tau_R}(x) &\simeq \frac{1}{\tau} \l[\langle q \rangle - x \r]\,, \\
  D^{(2)}_{\tau\gg\tau_R}(x) &\simeq \frac{1}{2\tau} \l[\langle q^2 \rangle - 2\langle q\rangle x + x^2 \r]\,.
\end{align}
\end{subequations}

That means that every Langevin process, independent of the true KM coefficients, appears as a process with linear drift and quadratic diffusion if the sampling interval of the data set is large compared to the relaxation time of the process. 
Therefore, as a preinvestigation before a KM analysis, 
it is necessary
to investigate the autocorrelation function of a process in order to estimate the relaxation time, and to estimate the conditional moments for various available $\tau$ and present it as a plot like Figures \ref{fig:OU_exmpl_1} and \ref{fig:OU_exmpl_2}. Furthermore, Ref. \cite{riera2010jsm} provides some checks in order to validate models with linear drift and quadratic diffusion.

We can conclude that in the regime $\tau\ll\tau_R$, finite time effects are small and low order corrections should suffice or can even be neglected. In the regime $\tau\gg\tau_R$, all information about the true dynamics is lost and a KM analysis is no longer possible. In between those two regimes, if $\tau\approx\tau_R$, a KM analysis is possible with use of our method presented in Ref. \cite{honisch2011pre}, as is demonstrated in several examples. Of course, at some point depending on the available amount of data, the results obtained by this method will also fail when $\tau$ approaches the limiting case of statistical independence. The point at which it fails will depend on the size of the available data set. The lower statistical uncertainties are in the estimation of the conditional moments, the more accurate one can extrapolate to $\tau=0$. To quantify this possible accuracy, we extend our method by the computation of errors in the estimated parameters with which drift and diffusion coefficients are parametrized. This is done by a Monte Carlo error propagation technique that is often used in nonlinear optimization problems \cite{aster2005}. 

In the next section, we first summarize the general method and present some minor improvements. Afterwards the Monte Carlo error propagation technique will be introduced in some more detail. We also address the limitations of the method due to finite amount of data and large sampling intervals.

\section{KM analysis at low sampling rates} \label{sec:method}

\subsection{Review of the method and minor improvements}

The optimization procedure introduced in \cite{honisch2011pre} is based on an inversion of Lade's method \cite{lade2009pla} to compute finite time coefficients with help of the AFPE, which was already mentioned in the foregoing section. The basic steps that have to be performed are the following. 

As a first step,  one has to obtain estimations of the conditional moments $\hat M^{(1,2)}_{\tau}(x)$ for several values of $x$ and $\tau$ from the time series data. For this we recommend a kernel based regression with use of the Nadaraya-Watson estimator. The bandwidth was in \cite{honisch2011pre} selected according to Silverman's rule of thumb \cite{haerdle2004} as
\begin{equation} \label{eq:silverman}
  h=1.06\hat{\sigma} N^{-1/5}\,,
\end{equation}
where $\hat{\sigma}$ is the standard deviation of the time series data. We found that this bandwidth is in many cases a bit too large leading to underestimated slopes of straight lines or curvatures of parabolas. More sophisticated data driven bandwidth selectors such as cross validation techniques or the method described in \cite{lamouroux2009pla} are computationally expensive and did not yield robust improvements in our tested cases. We obtained the best results by simply reducing the factor 1.06 in \eqref{eq:silverman} to 0.8.

The next step is to find a suitable parametrization for the drift and diffusion coefficients. In most cases a polynomial ansatz will be appropriate. Also a representation in form of spline curves has been successfully tested in \cite{honisch2011pre}, where the interpolation points serve as optimization parameters. In either case, the coefficients are represented as $\KM {1,2}(x,\sigma)$ with a set of optimization parameters $\sigma$. If an analytic expression has been chosen for the parametrization, one can try to obtain an analytic expression for the corresponding conditional moments $M^{(1,2)}_{\tau}(x,\sigma)$ via Lade's method. Otherwise, they have to be calculated by numerical integration of the AFPE. The optimal set of parameters must then be obtained by minimizing a distance measure between the estimated conditional moments $\hat M^{(1,2)}_{\tau}(x)$ and the computed ones. In \cite{honisch2011pre} we have chosen the weighted least squares distance
\begin{eqnarray}
  V(\sigma) = \sum_{i=1}^M \sum_{j=1}^N \l[ \frac{\l\{ \hat M^{(1)}_{\tau_i}(x_j) -  M^{(1)}_{\tau_i}(x_j,\sigma) \r\}^2}{\l(\hat \sigma_{ij}^{(1)} \r)^2  } \r. \nonumber\\+ \l.\frac{\l\{ \hat M^{(2)}_{\tau_i}(x_j) -  M^{(2)}_{\tau_i}(x_j,\sigma) \r\}^2}{\l(\hat \sigma_{ij}^{(2)} \r)^2  } \r]~, \label{eq:pot}
\end{eqnarray}
where $\hat \sigma_{ij}^{(1,2)}$ are the statistical errors in the estimation of the conditional moments. In \cite{honisch2011pre} we used 
\begin{subequations}
\label{eq:errors_old}
\begin{align}
  \hat \sigma_{ij}^{(1)} &= \sqrt{ \frac{\hat M^{(2)}_{\tau_i}(x_j) - \l( \hat M^{(1)}_{\tau_i}(x_j) \r)^2}{ \sum_{k=1}^T \frac 1h K\l(\frac{x_j-X_{t_k}}{h} \r) }}  \\
  \hat \sigma_{ij}^{(2)} &= \sqrt{ \frac{\hat M^{(4)}_{\tau_i}(x_j) - \l( \hat M^{(2)}_{\tau_i}(x_j) \r)^2}{ \sum_{k=1}^T \frac 1h K\l(\frac{x_j-X_{t_k}}{h} \r) }}
\end{align}
\end{subequations}
as an error estimate, where $T$ is the number of data points of the time series data, $K(\bullet)$ is the kernel function, and $h$ is the selected bandwidth. The error estimate \eqref{eq:errors_old} does not take into account the influence of the bandwidth. According to  \cite{haerdle2004}, Eqs. \eqref{eq:errors_old} have to be corrected by a factor of 
\begin{equation} \label{eq:corr_fac}
  \sqrt{\frac{||K||_2}{h}}~,
\end{equation}
where $||K||_2^2$ is the $L_2$ norm of the kernel, thus
\begin{equation}
  ||K||_2 = \sqrt{ \int K^2(x) \md x }~.
\end{equation}
For the Epanechnikov kernel
\begin{equation} \label{eq:epanechnikov}
  K(x) = \begin{cases}
                    \frac{3\sqrt{5}}{100} \l(5-x^2\r)\quad &\text{if}~x^2<5 \\
                    0                                  &\text{if}~x^2>5
                  \end{cases}~,
\end{equation}
which we use, $||K||_2\approx0.518$. However, regarding the optimization, the correction factor \eqref{eq:corr_fac} has no influence, since the error estimate has only the purpose of a relative weighting of the estimated conditional moments for different values of $x$ and $\tau$. It only becomes relevant in connection with the Monte Carlo error propagation described in Sec. \ref{ssec:MCEP}.

The minimum of \eqref{eq:pot} is in \cite{honisch2011pre} determined by a trust region algorithm, but other optimization methods should work as well.

\subsection{Monte Carlo error propagation} \label{ssec:MCEP}

The basic idea of the Monte Carlo error propagation (MCEP) approach is as follows: After estimating the model parameters of a specific model from a noisy data set (backward problem), the model is used to produce an unnoisy data set (forward problem). Then one generates an ensemble of pseudo data sets by perturbing the unnoisy data set with different realizations of noise. For each data set again, the model  parameters are estimated. Eventually, one can compute the standard deviation for each model parameter from this ensemble, which can be used as an uncertainty measure.

In the present case, we do not regard the time series data as our noisy data set, but the estimated conditional moments $\hat M^{(1,2)}_{\tau}(x)$. In order to obtain reliable uncertainty estimates with the MCEP approach, it is crucial that the artificial perturbations of the unnoisy data set are of the same amplitude as the errors of the original data set. Therefore one needs a reliable absolute error estimate for the conditional moments and the correction factor \eqref{eq:corr_fac} becomes important.

To generate the ensemble of pseudo data sets, one has to compute the conditional moments that correspond to the estimated model parameters by solving the corresponding AFPE. To each $M^{(1,2)}_{\tau_i}(x_j)$ we add an independent Gaussian distributed random number with standard deviation of expression \eqref{eq:errors_old} multiplied by the factor \eqref{eq:corr_fac} for each member of the data set ensemble. Then the model parameters are estimated again for each data set and the corresponding standard deviation is calculated for each model parameter.

There is one technical problem connected to the MCEP approach. It can happen that the optimization routine does not find the absolute minimum of the least square potential for some pseudo data sets of the ensemble. Although the trust region algorithm is a very robust one, our observation is that it sometimes gets stuck in local minima close to the initial condition. If the initial condition is selected the same for all pseudo data sets, it can therefore happen that the computed standard deviation is significantly underestimated. To overcome this problem, we choose random initial conditions that lie in a region around the estimated parameters from the original data set. This in return bears the danger that the optimization routine will not find the way back to the true minimum if the initial condition lies too far away. This can lead to overestimated standard deviations. Therefore, only optimization results are accepted where the final residual is below a specific threshold. Otherwise the optimization is repeated with another initial condition. The threshold is selected three times the residual of the result from the original data set. Nevertheless it can happen that some outliers lead to a significantly overestimated standard deviation, especially in cases of very large sampling intervals. The easiest way to overcome this problem is to detect these outliers by eye and remove them from the ensemble.

\begin{figure}[htb]
  \includegraphics[width=.47\textwidth]{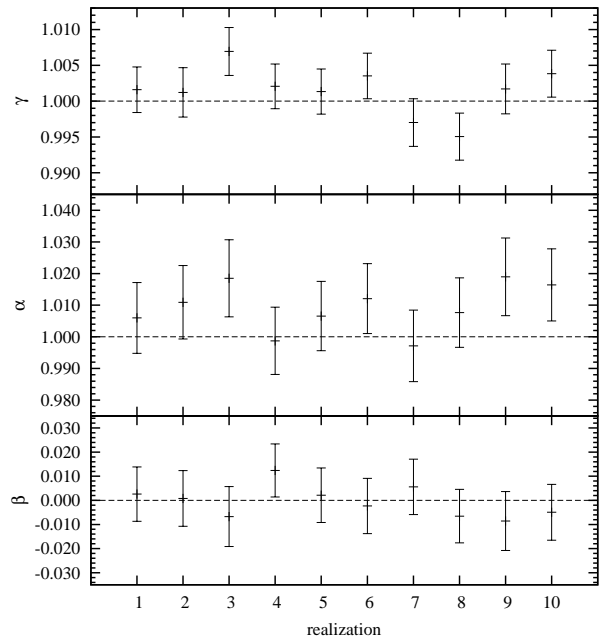}
  \caption{Estimated parameters for an ensemble of ten OU processes according to Eqs. \eqref{eq:lin_quad} with different noise realizations. The error bars are computed via the MCEP method. The horizontal dashed lines indicate the true values of the parameters. The sampling interval $\tau$ of the series is equal to the relaxation time $\tau_R=1$.} \label{fig:MCEP_test}
\end{figure}

\begin{table}[htb]
  \caption{Comparison between the standard deviations $\sigma_E$ of estimated parameters in the ensemble of ten OU processes and the mean errors $\overline{\sigma_{MCEP}}$ computed by the MCEP approach.} \label{tab:MCEP_test}
  \begin{ruledtabular}
    \begin{tabular}{cccc}
      Parameter& average & $\sigma_E$ & $\overline{\sigma_{MCEP}}$ \\
      \hline
      $\gamma$ & ~1.0014 & 0.0035 & 0.0033 \\
      $\alpha$ & ~1.009~ & 0.012~ & 0.014~ \\
      $\beta$  & -0.0006 & 0.0063 & 0.0116 \\
    \end{tabular}
  \end{ruledtabular}
\end{table}

To test the MCEP approach, we generate an ensemble of ten synthetic time series of an OU process with $\KM 1 = -x$ and $\KM 2 = 1$. The corresponding Langevin equation is integrated with the Euler-Maruyama scheme with a time increment $\Delta t=10^{-2}$. Only every 100th data point is stored so that the sampling interval of time series is $\tau=1$. Each time series consists of $10^6$ data points. For the optimization we choose the parametrization
\begin{subequations} \label{eq:lin_quad}
\begin{align}
  \KM1(x) &= -\gamma x\,, \\
  \KM2(x) &= \alpha + \beta x^2 \,, \label{eq:quad_diff}
\end{align} 
\end{subequations}
with the optimization parameters $\gamma,\alpha$ and $\beta$. For each of the ten data sets, we estimate these parameters as well as the corresponding standard deviations with the MCEP approach. The results are depicted in Fig. \ref{fig:MCEP_test}. By eye, the estimated error bars seem reasonable. To make a quantitative statement, we compute the standard deviations of the estimated parameters from the ensemble and compare them to the mean standard deviations of the MCEP approach. The numbers are listed in Table \ref{tab:MCEP_test}. The errors of the parameters $\gamma$ and $\alpha$ fit very well while the error of $\beta$ is a bit overestimated.

However, one should note that the errors obtained by the MCEP approach only cover uncertainties caused by statistical fluctuations due to the finite amount of data. The influence of other error sources such as measurement noise, deviations from the Markov property, non-stationarity of the time series or an inappropriate parametrization of drift and diffusion coefficients, just to name a few, are not considered. Therefore, the MCEP errors should be regarded as a lower bound for the true errors. Nevertheless they can decrease the danger of an overestimation of the significance of KM analysis results.

\subsection{Limitations of the approach} \label{sec:limitations}

Now we use the MCEP approach to demonstrate the limitations of the KM analysis caused by finite time effects and limited amount of data. To this end, we compute the error estimates for the parameters $\gamma$, $\alpha$ and $\beta$ for synthetic time series with different sampling intervals. The synthetic time series are generated in the same manner as described in section \ref{ssec:MCEP}. All data sets consist of $10^6$ data points. 

\begin{figure}[htb]
  \includegraphics[width=0.47\textwidth]{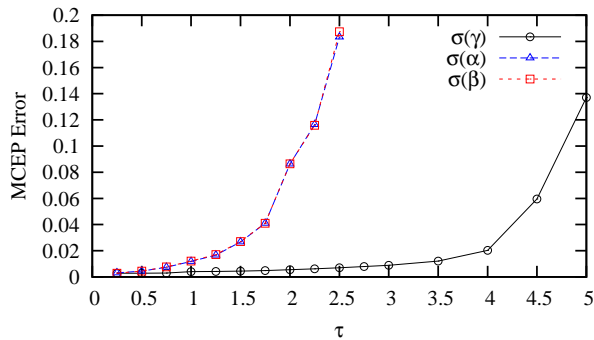}
  \caption{(Color online) Estimated MCEP errors for synthetic time series according to the model \eqref{eq:lin_quad} with parameters $\gamma=\alpha=1$ and $\beta=0$, as a function of the sampling interval $\tau$.} \label{fig:error_tau}
\end{figure}

The result is shown in Fig. \ref{fig:error_tau}. One can see that the estimation of the diffusion parameters $\alpha$ and $\beta$ breaks down around 2.5 relaxation times ($\tau_R=1$), while the drift parameter $\gamma$ can still be estimated up to about 5 relaxation times. Above these values it is difficult to obtain robust error estimates with the MCEP approach because the afore-mentioned outliers become very frequent. However, using the MCEP approach for sampling intervals above these values, one will clearly notice that the KM analysis is no longer feasible.

Looking at Fig. \ref{fig:error_tau}, one further notices that the errors of $\alpha$ and $\beta$ are strongly correlated, which is not surprising. That the absolute sizes of the errors are almost equal is only the case for the specific selection of parameters $\gamma=\alpha=1$ and $\beta=0$.

\section{Application to an optical trapping experiment} \label{sec:application}

\subsection{Experimental setup} \label{sec:setup}

\begin{figure}
  \centering
  \includegraphics[width=1.0\columnwidth]{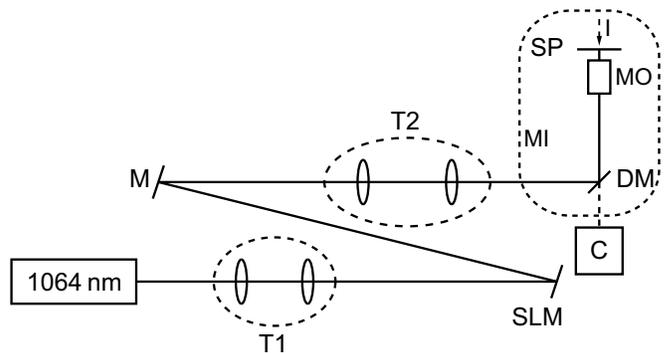}
  \caption{Experimental setup: (D)M: (dichroic) mirror; C: CMOS camera; I: illumination; MI: microscope; MO: microscope objective; T1 and T2: telescope systems to achieve an optimal beam diameter; SLM: phase-only spatial light modulator, for the described experiment not connected to a power supply and thus acting as a mirror; SP: sample plane, where \unit{20}{\micro\liter} of the used suspension was fixed between a microscope slide and a coverglass.}
  \label{setup}
\end{figure}

Trapping of small dielectric objects in the nano- and microscale can be realized by focusing a single laser beam so that it creates a gradient force near the focal region that is able to hold these particles, and even levitate them against gravity \cite{ashkin1980science}. Our trapping experiment is based on such a single beam gradient laser tweezers system (see FIG. \ref{setup}) as in \cite{hoerner2010jbp}.
A near infrared laser ($\lambda=\unit{1064}{\nano\metre}$, maximum output power of \unit{2.5}{\watt}, Smart Laser Systems) with a Gaussian beam distribution is coupled into an inverted  microscope (Ti Eclipse, Nikon) for trapping with an oil-immersion objective (CFI Apo TIRF 100x) with a numerical aperture of $N\!A=1.49$. This objective is also used for observation purposes and images the observation plane onto a high-speed CMOS sensor (MV2-D1280-640-CL-8, Photonfocus) with a frame rate of \unit{488}{\hertz} at full resolution. By reducing the size of the captured images to a region of interest of 80\,x\,64 pixels, the frame\-rate could be increased to \unit{3966}{\hertz}, which corresponds to a sampling interval $\tau_s\approx \unit{0.25}{\milli \second}$. For the measurements uncoated polystyrene beads (Kisker) with a diameter of \unit{(1.002\pm0.043)}{\micro\metre} are used. These microparticles are suspended in destilled water and trapped 5--\unit{10}{\micro\metre} above the cover glass. Due to technical reasons up to 150 videos consisting of $10^4$ frames are captured over 15 minutes and analyzed afterwards. The particle's position is determined by a Matlab algorithm implementing a center of mass detection \cite{Crocker1996,Blair}. 

The autocorrelation function of a trapped particle's x(y)-position decays with a relaxation time $\tau_{x(y)}= 6\pi\eta r\per k_{x(y)}$  where $\eta$ is the viscosity of the surrounding fluid, $r$ the radius of the particle and $k_{x(y)}$ the corresponding trap stiffness \cite{Meiners1999}. For a particle with a radius of $r=\unit{0.5}{\micro\metre}$ typical relaxation times are around \unit{10}{\milli\second \, \micro\metre \, \pico\newton^{-1}}\per$k_x$. Therefore the laser power is reduced to about \unit{10}{\milli\watt} in the trapping plane which results in a trapping stiffness of $k_x=\unit{21}{\pico\newton\per\micro\metre}$ and thus a relaxation time of about $\tau_x=\unit{0.45}{\milli\second}$ (compare Fig. \ref{fig:ac}) in order to have sufficient temporal resolution to observe the particle's dynamics.

\subsection{Modeling} \label{sec:modeling}

The equation of motion for a Brownian particle in an optical trap reads for one spatial dimension \cite{bedeaux1974physica}
\begin{equation} \label{eq:BM}
  m_p \ddot x(t) = F_{\mathrm{fr}}(t) + F_{\mathrm{op}}(x(t)) + F_{\mathrm{th}}(t)\,,
\end{equation}
where $m_p$ is the mass of the particle, $F_{\mathrm{fr}}$ is the friction force, $F_{\mathrm{op}}$ is the force induced by the optical trap and $F_{\mathrm{th}}$ denotes the thermal fluctuations. The generell form of the friction force reads
\begin{equation} \label{eq:ret_fric}
  F_{\mathrm{fr}}(t) = - \int \gamma(t-t') \dot x(t') \md t'\,.
\end{equation}
According to the fluctuation dissipation theorem \cite{kubo1991book}, the kernel $\gamma(t)$ is connected to the correlation function of the thermal fluctuations
\begin{equation} \label{eq:fluc_diss}
  \langle F_{\mathrm{th}}(t) F_{\mathrm{th}}(t') \rangle = k_B T \gamma(|t-t'|)\,.
\end{equation}
Assuming a laminar velocity profile around a spherical particle and no-slip boundary conditions, the kernel is given by
\begin{equation}
  \gamma(t-t') =  2\delta(t-t') 6\pi \eta r 
\end{equation}
where $\eta$ is the dynamic viscosity of the fluid and $r$ the radius of the particle. This yields the well-known Stokes' law
\begin{equation}
  F_{\mathrm{fr}}(t) = -\lambda \dot x(t)
\end{equation}
with $\lambda=6\pi \eta r$.
On time scales $\tau\gg \tau_I = m_p/\lambda$ $(\tau_I \sim \unit{0.05}{\micro \second}$ in our experimental setup), inertia can be neglected. With a linear optical force $F_{\mathrm{op}}=-kx$, this yields
\begin{equation} \label{eq:langevin_op}
  \dot x(t) = -\frac{k}{\lambda}x(t) + \sqrt{2\KM 2} \varGamma(t)\,,
\end{equation}
with $\langle \varGamma(t)\varGamma(t') \rangle=\delta(t-t')$ and the diffusion coefficient
\begin{equation} 
  \KM 2 = \frac{k_B T}{6 \pi \eta(T) r}\,, \label{eq:Einstein_Stokes}
\end{equation}
which is known as the Einstein-Stokes equation. For a constant temperature equal to the room temperature of $T=(294\pm2)\,$K, a particle diameter of $2r=\unit{(1.002\pm0.043)}{\micro\metre}$, and a viscosity of $\eta=$(1.00$\pm$0.02)$\cdot10^{-3}$\,\unit{}{\newton \second \metre^{-2}}, the Einstein-Stokes equation predicts a diffusion constant of 
\begin{equation} \label{eq:diff_ES}
    \KM 2_{\text{ES}} = \unit{(0.43\pm0.03)}{(\micro \metre)^2\second^{-1}}\,.
\end{equation}

A more realistic treatment goes beyond Stokes' law and also considers the momentum that is transferred from the particle to the fluid. Ref. \cite{bedeaux1974physica} gives a derivation of Eqs. \eqref{eq:BM}, \eqref{eq:ret_fric}, and \eqref{eq:fluc_diss} for a macroscopic sphere in an incompressible fluid from linearized stochastic hydrodynamic equations. Thereby the Fourier transform of the memory kernel is computed as
\begin{equation}
  \hat{\gamma}(\omega) = 6\pi \eta r \l[1 + (1-\mi)r\sqrt{\frac{\omega \rho_f}{2\eta} } - \frac{\mi \omega \rho_f r^2}{9\eta} \r]\,.
\end{equation}
Here, $\rho_f$ denotes the mass density of the fluid. This corresponds to a friction force
\begin{align}
  F_{\mathrm{fr}}(t) = &-6\pi \eta r \dot x(t) - \frac{m_f}{2} \ddot x(t) \nonumber \\
  &- 6 r^2 \sqrt{\pi \rho_f \eta} \int_{-\infty}^t \md t' \frac{\ddot x(t')}{\sqrt{t-t'}}\,, \label{eq:basset}
\end{align}
where $m_f=(4\pi/3)\rho_f r^3$ is the mass of the displaced fluid. The correlation of the thermal fluctuations shows a negative algebraically decaying tail
\begin{equation}
  \langle F_{\mathrm{th}}(t) F_{\mathrm{th}}(t') \rangle = -3 r^2 k_B T \sqrt{\pi \rho_f \eta} |t-t'|^{-3/2} \label{eq:cor_fluc}
\end{equation}
for $|t-t'|^{-3/2} > 0$. The memory term in the friction force \eqref{eq:basset} and the long correlations of the thermal fluctuations clearly obliterate the Markov property of the process. 
However, it is possible that a finite Markov-Einstein (ME) time scale $\tau_{\text{ME}}$ exists, which means that the process can approximately be described by a Markov process for times larger than $\tau_{\text{ME}}$. 
If such a time scale is found, one can try to find a model in form of Eq. \eqref{eq:langevin_op} with an effective drag coefficient $\lambda$ and diffusion coefficient $\KM 2$ with our data analysis method.

\begin{figure}[htb]
  \includegraphics[width=.47\textwidth]{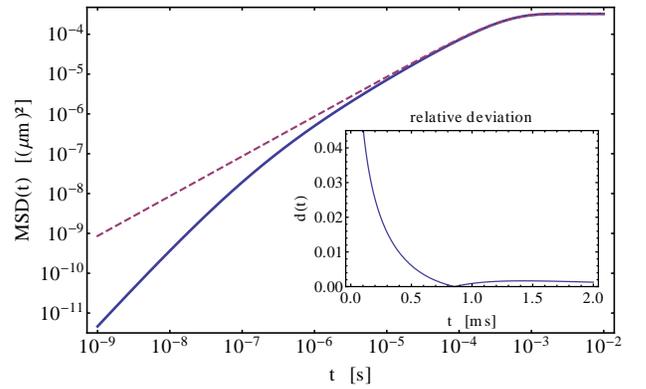}
  \caption{(Color online) Mean squared displacement against time according to the full hydrodynamic model (blue solid line) and the overdamped Markov model (red dashed line) as a double logarithmic plot. The inset shows the relative deviation, Eq. \eqref{eq:rel_dev}, between both curves.} \label{fig:msd_bm_pot}
\end{figure}

To obtain a rough estimation of the ME time scale, we compare the mean squared displacement $MSD(t)=\langle (\Delta x(t) )^2\rangle$
of the full hydrodynamic model, which was computed by Clercx \cite{clercx1992pra}, with the MSD of the Markov model, Eq. \eqref{eq:langevin_op}.  Fig. \ref{fig:msd_bm_pot} shows a plot of the MSDs for both models with the parameters according to our experiment. In the overdamped Markov model (red dashed curve), there is only one characteristic time scale $\tau_k=\lambda/k\sim \unit{10^{-4}}{\second}$ depending on the stiffness $k$ of the optical trap. Below this time scale there is the so called diffusive regime in which the MSD grows linear with time, $MSD_M(t\ll\tau_k)=2\KM2 t$. Above this time scale the MSD saturates to the constant value $2k_B T/k$.

In the full hydrodynamic model the smallest characteristic time scale is the inertia time scale $\tau_I\sim \unit{10^{-8}}{\second}$. For times smaller than $\tau_I$ the MSD grows quadratically with time, $MSD_H(t\ll\tau_I) = (k_B T/m^*)t^2$, where $m^*=m_p + m_f/2$. This is the so called ballistic regime \footnote{Below this regime there is the additional characteristic time scale of sound propagation in which the incompressibility assumption becomes invalid. But this is not covered by the full hydrodynamic model.}. Above the ballistic regime it takes about four decades until the MSD approaches the diffusive regime and coincides with the overdamped Markov model. These four decades are influenced by hydrodynamic memory effects.
The inset of Fig. \ref{fig:msd_bm_pot} shows the relative deviation
\begin{equation} \label{eq:rel_dev}
  d(t) = \frac{|MSD_H(t) - MSD_M(t)|}{MSD_H(t)}
\end{equation}
between the MSDs of the two models. According to this, the influence of the hydrodynamic memory effects is still present at times $t\sim\tau_k$.
For times larger than \unit{0.4}{\milli \second}, the relative deviation becomes smaller than one percent. Therefore, we expect a ME time scale of this order of magnitude.
A direct test of our data for Markovianity is presented in Sec. \ref{sec:preinvest} yielding a ME time scale of $\tau_{\text{ME}} \approx \unit{0.5}{\milli \second}$ which is in good agreement to the discussion above.

\subsection{Data preparation}

\begin{figure}[htb] 
  \includegraphics[width=0.47\textwidth]{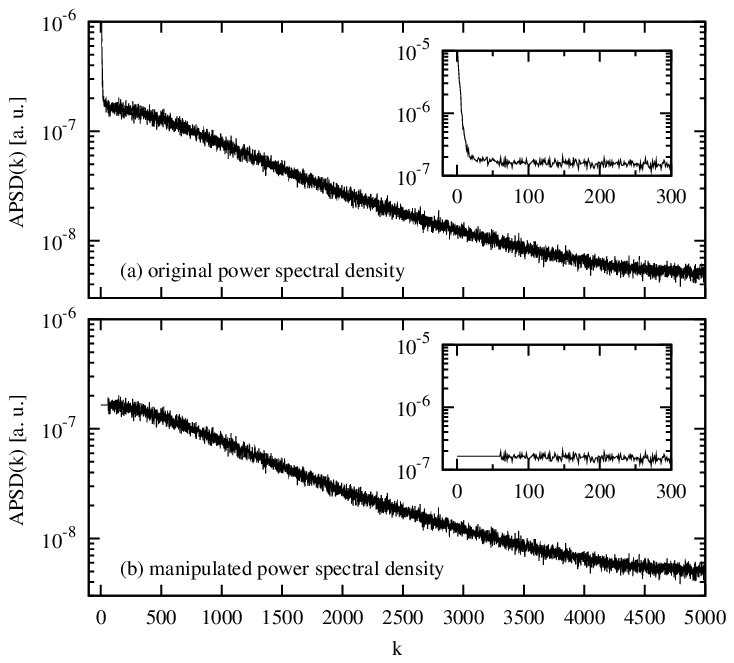}
  \includegraphics[width=0.47\textwidth]{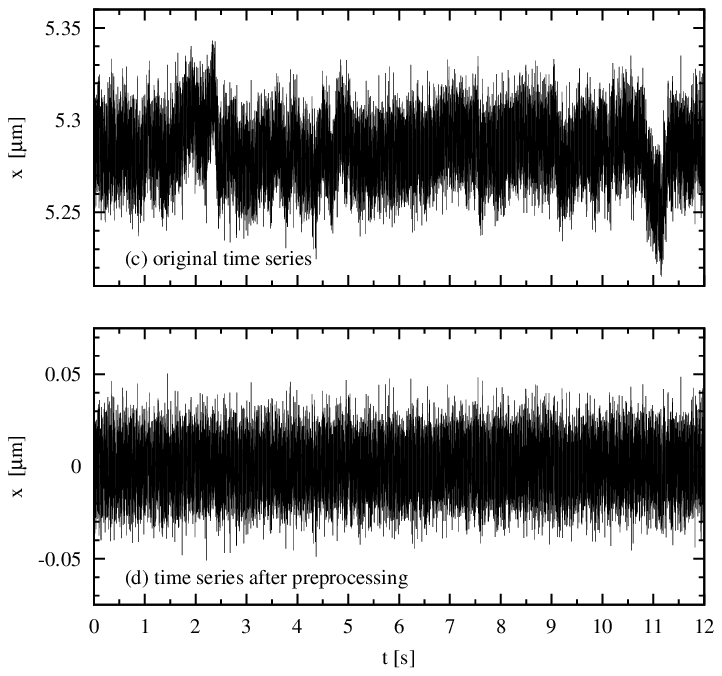}
\caption{Preprocessing of the data set. Upper panels: Averaged power spectral density (APSD) of the $x$ coordinate of the particle motion before (a) and after (b) preprocessing in arbitrary units. The insets show enlargements of the low wave number part, where the filtering occurs. Lower panels: Excerpt of the corresponding time series before (c) and after (d) preprocessing.} \label{fig:detrending}
\end{figure}

Fig. \ref{fig:detrending} (c) shows the first twelve seconds of the $x$ component of the original data set. One notices that the mean value of the time series fluctuates over time, i.\,e., the time series is not stationary. These fluctuations of the mean value lead to large values in the power spectral density (PSD) of the time series at low wave numbers $k$ as one can see in Fig. \ref{fig:detrending} (a), which shows the PSD averaged over the 150 measured trajectories. To enhance the stationarity of the time series, the corresponding Fourier coefficients for each of the 150 trajectories are reduced such that they fit in the Lorentz-like form of the spectrum. The zero wave number coefficients are set to zero to center time series around zero. This kind of high pass filtering leads to the averaged power spectral density depicted in Fig. \ref{fig:detrending} (b). The corresponding time series after this preprocessing is shown in Fig. \ref{fig:detrending} (d). The same type of preprocessing is applied to the $y$ component of the trajectories.

\subsection{Preinvestigations} \label{sec:preinvest}

\begin{figure}[b]
  \includegraphics[width=.47\textwidth]{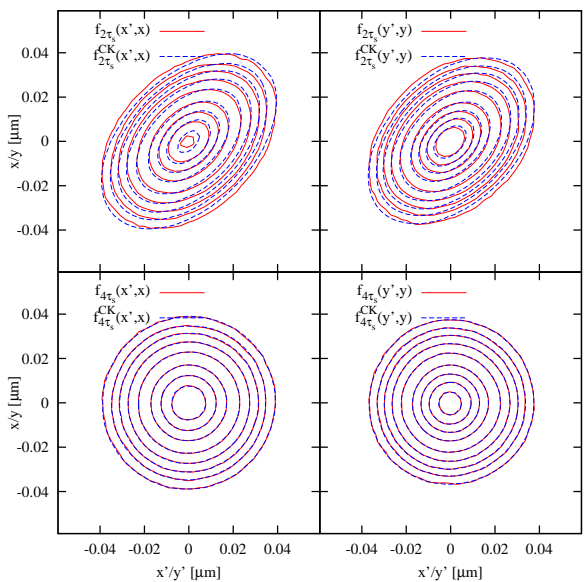}
  \caption{(Color online) Test of the validity of the CKE for a time scale equal to the sampling interval (upper panels) and twice the sampling interval (lower panels) of the $x$ component (left panels) and $y$ component (right panels) of the data set. The plots show contour lines of estimated joint PDFs, according to Eqs. \eqref{eq:pdf_def}. In the lower panels the contour lines match well in contrast to the upper panels indicating a ME time scale of $\tau_{\mathrm{ME}} \approx 2\tau_s$.} \label{fig:martest}
\end{figure}

As a first preinvestigation, we take a look at the Markov property. A necessary condition for a process to be Markovian on a specific time scale $\tau$ is the validity of the Chapman-Kolmogorov equation (CKE) \cite{risken}:
\begin{equation}
 p_{2\tau}(x'|x) = \int \md x'' p_{\tau}(x'|x'') p_{\tau}(x''|x)\,,
\end{equation}
where $p_{\tau}(x'|x):=p(x',t+\tau|x,t)$ denotes the transition probability density function of the process. We further define 
\begin{subequations} \label{eq:pdf_def}
\begin{align}
  p_{\tau}^{\mathrm{CK}}(x'|x) &:= \int \md x'' p_{\tau/2}(x'|x'') p_{\tau/2}(x''|x)\,, \\
  f_{\tau}^{\cdots}(x',x) &:= p_{\tau}^{\cdots}(x'|x) f(x)\,,
\end{align}
\end{subequations}
To test our data for Markovianity on a time scale $\tau$, we compare estimates for the joint PDFs $f_{2\tau}$ and $f_{2\tau}^{\mathrm{CK}}$ (cf. Ref. \cite{friedrich2011pr}). Fig. \ref{fig:martest} shows the corresponding contour plots for $\tau=\tau_s$ and $\tau=2\tau_s$ for the two components of the process. For $\tau=\tau_s$ (upper panels), one can see clear deviations between the contour lines which indicate that the process is not Markovian on this time scale. For $\tau=2\tau_s$ the deviations vanish.
This leads to the conclusion that the process has a ME time scale $\tau_{\mathrm{ME}} \approx 2\tau_s$. Therefore we only include conditional moments with time increments $\tau\geq2\tau_s$ into the minimization of Eq. \eqref{eq:pot}.

\begin{figure}[htb]
  \includegraphics[width=0.47\textwidth]{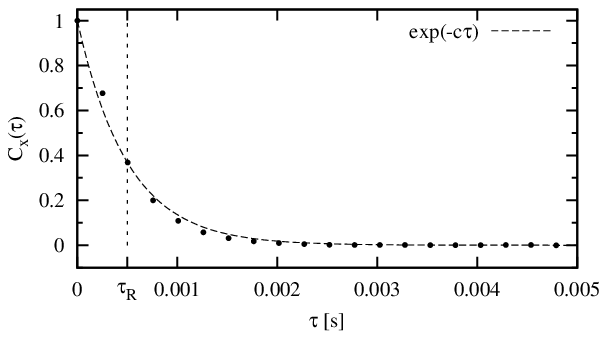} \\
  \includegraphics[width=0.47\textwidth]{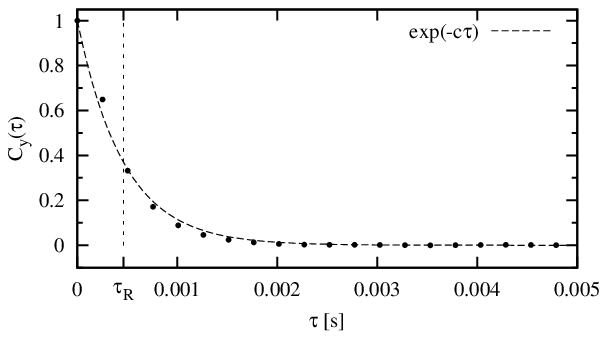}
  \caption{Autocorrelation functions for the $x$ (top) and $y$ (bottom) component of the particle motion. To obtain a rough estimation of the relaxation time $\tau_R$, we fit an exponential $\exp(-c\tau)$ to the first points of the correlation function. This leads to the estimate $\tau_R \approx c^{-1}$.} \label{fig:ac}
\end{figure}

As a next step, we take a look at the autocorrelation functions (ACF)
\begin{subequations} \label{eq:ACF}
\begin{align}
  C_x(\tau) &= \frac{\langle X(t) X(t+\tau) \rangle_t}{\langle (X(t))^2\rangle_t}\,, \\
  C_y(\tau) &= \frac{\langle Y(t) Y(t+\tau) \rangle_t}{\langle (Y(t))^2\rangle_t}
\end{align}
\end{subequations}
to get an impression of the typical time scales of the system and to decide whether the sampling interval $\tau_s$ is sufficiently small for a reliable KM analysis. The ACFs are shown in Fig. \ref{fig:ac}. 
To evaluate the relaxation time as a typical time scale of the system, we fit an exponential, $\me^{-c\tau}$, to the first points of ACF and take $\tau_R \approx c^{-1}$ as a rough estimate. According to this estimate, $\tau_R\approx2\tau_s$.
This means that the relaxation time of the process is approximately equal to the ME time scale.
Therefore, according to the discussion in Sec. \ref{sec:limitations}, the KM analysis should be possible.

\begin{figure}[bth]
  \includegraphics[width=0.47\textwidth]{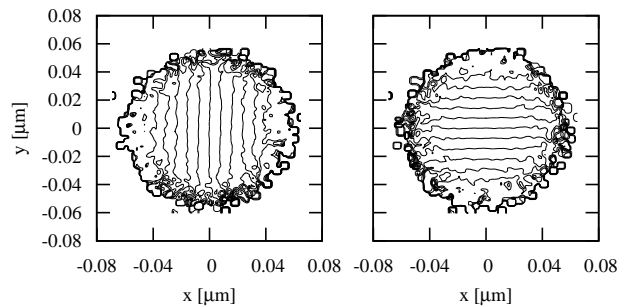}
  \caption{Contour plot of the $x$ component (left) and the $y$ component (right) of the measured drift vector field. As one can see, the $x$ component does not significantly depend on $y$ and vice versa.} \label{fig:drift_contour}
\end{figure}

As a last step, we check whether it is possible or not to regard the $x$ and $y$ components of the particle motion as two independent processes. To this end, we measure the finite time drift vector field $\bs D^{(1)}(x,y,\tau) = (\KM 1_x(x,y,\tau) , \KM 1_y(x,y,\tau))^{\top}$ with
\begin{align*}
  \KM 1_x(x,y,\tau) &= \frac{1}{\tau} \langle X(t+\tau) - X(t) | X(t)=x; Y(t)=y \rangle \\
  \KM 1_y(x,y,\tau) &= \frac{1}{\tau} \langle Y(t+\tau) - Y(t) | X(t)=x; Y(t)=y \rangle
\end{align*}
for $\tau=\tau_s$. We assume that the qualitative form of the drift vector field is not affected that much from finite time effects and deviations from the Markov property. In Fig. \ref{fig:drift_contour} we show contour plots of $\KM 1_x(x,y,\tau)$ (left) and $\KM 1_y(x,y,\tau)$ (right). In a region with a radius of about \unit{0.04}{\micro\metre} around the origin of the coordinate system, where most of the data points are, the contour lines of $\KM 1_x(x,y,\tau)$ and $\KM 1_y(x,y,\tau)$ are approximately linear and parallel to the $y$ axis and $x$ axis, respectively. Hence,
\begin{subequations}
\begin{align}
    \KM 1_x(x,y,\tau) &= \KM 1_x(x,\tau)\,, \\
    \KM 1_y(x,y,\tau) &= \KM 1_y(y,\tau)\,.
\end{align}
\end{subequations}
Therefore, we treat the two components of the particle motion as independent processes.

\subsection{Analysis results}

To estimate the drift and diffusion coefficients for each of the two processes, we first make a parametric ansatz in form of an OU process, Eqs. \eqref{eq:OU}, with optimization parameters $\gamma$ and $\alpha$. Our experience from synthetic time series data is that the best results are achieved if one includes the finite time coefficients of the $n$ smallest time increments $\tau_i=i\tau_s$, such that $\tau_n$ is between one and two relaxation times. Since we have a finite ME time scale of $2\tau_s$  in our case, we include the conditional moments with $\tau=i\tau_s$, $i=2,3,4$ into the least squares potential, Eq. \eqref{eq:pot}.

\begin{table}[htb]
  \caption{Results for the optimization parameters $\gamma$ and $\alpha$ for the $x$ and $y$ components of the process together with the error estimates $\sigma_{MCEP}$ obtained by the MCEP method.} \label{tab:results}
  \begin{ruledtabular}
    \begin{tabular}{ldd}
	& \multicolumn{1}{l}{$\gamma~[\mathrm{s}^{-1}] $}& \multicolumn{1}{l}{$\alpha~[(\unit{}{\micro \metre})^2 \mathrm{s} ^{-1}]$} \\
     \hline
     $x$ comp. & & \\
     \hline
     result & 2004.6 & 0.32910  \\
     $\sigma_{MCEP}$ & 4.2 & 0.00064 \\
     \hline
     $y$ comp. & & \\
     \hline
     result & 2212.4 & 0.32240 \\
     $\sigma_{MCEP}$ & 4.8 & 0.00067 \\
    \end{tabular}
  \end{ruledtabular}
\end{table}

\begin{figure}[t!]
  \includegraphics[width=.47\textwidth]{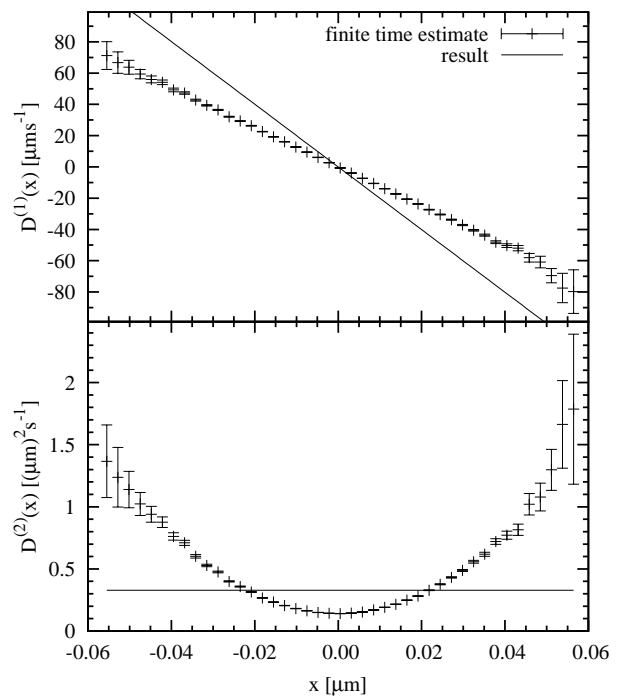}
  \caption{Result of the optimization for drift (top) and diffusion (bottom) coefficients of the $x$ coordinate of the particle motion. The symbols with the error bars are the estimated finite time coefficients for the smallest available time increment above the ME time scale, i.\,e. $\tau=2\tau_s$. The optimized coefficients are represented by the solid lines.} \label{fig:res_x}
\end{figure}

\begin{figure}[t!]
  \includegraphics[width=.47\textwidth]{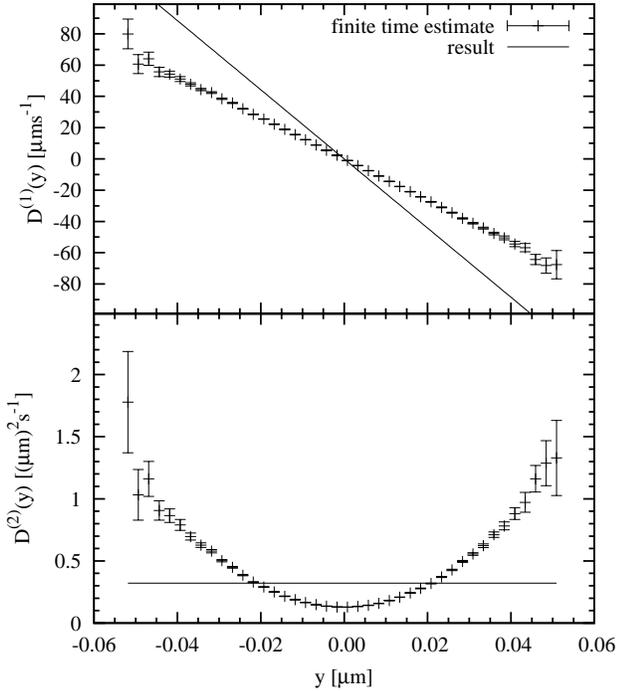}
  \caption{Result of the optimization for drift (top) and diffusion (bottom) coefficients of the $y$ coordinate of the particle motion. The representation is analogous to Fig. \ref{fig:res_x}.} \label{fig:res_y}
\end{figure}

Table \ref{tab:results} shows the results of the optimization as well as the estimated errors by the MCEP method for the $x$ and $y$ components of the process. A graphical representation is depicted in Figures \ref{fig:res_x} and \ref{fig:res_y}, respectively.
A possible spatial dependence of the temperature due to the heating of the laser cannot be resolved on the basis of the experimental data. If one includes a quadratic term in the diffusion ansatz as in Eq. \eqref{eq:quad_diff}, the error estimate for the parameter $\beta$ is of the same size as the estimated value. However, our model with linear drift and constant diffusion describes the process very well, as we will see in Sec. \ref{sec:comparison}.

The diffusion coefficients deviate by a factor of approximately 1.3 from the result that was expected from the Stokes-Einstein equation (s. Sec. \ref{sec:modeling}). To understand this deviation, we also measure the diffusion coefficients of different freely diffusing particles. The results are presented in the following section.
The stiffness of the trap can nevertheless be calculated by
\begin{equation}
  k = k_B T\frac{ \gamma}{\alpha}\,.
\end{equation}
If we assume a temperature of (294$\pm$2)\,K, we obtain
\begin{align}
  k_x &= (24.72\pm0.27) \frac{\text{pN}}{\unit{}{\micro\metre}}\,, \\
  k_y &= (27.85\pm0.31) \frac{\text{pN}}{\unit{}{\micro\metre}}\,.
\end{align}

\subsection{Diffusion coefficient of freely diffusing particles}

\begin{figure}[htb]
  \includegraphics[width=.47\textwidth]{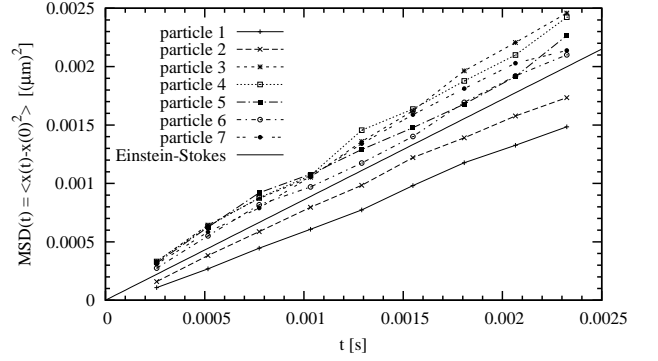}
  \caption{Mean squared displacement of the $x$ coordinate as a function of time for seven freely diffusing particles. The diffusion coefficient can be determined by a linear fit for each particle. The solid straight line shows the MSD that is expected according to the Einstein-Stokes equation, Eq. \eqref{eq:diff_ES}.} \label{fig:msd}
\end{figure}

In order to understand the deviations from the Einstein-Stokes equation, we also determine the diffusion coefficients for freely diffusing particles. In the same manner as described in Sec. \ref{sec:setup}, we measure the positions of seven different particles that are not trapped by optical tweezers. For each particle, one time series with approximately $10^4$ time steps at a sampling frequency of \unit{3873}{\hertz} is measured. From this data we determine the mean squared displacement (MSD) for which the relation
\begin{equation}
  \text{MSD}(t) := \langle (x(t) - x(0))^2 \rangle = 2 \KM 2 t
\end{equation}
holds. Fig. \ref{fig:msd} shows the obtained MSDs for the $x$ components of the seven particles. The diffusion coefficients can be determined by linear fits. If one averages the determined coefficients over all particles and both coordinate directions, one obtains
\begin{equation}
  \KM 2 = (0.44 \pm 0.06)\, \unit{}{(\micro\metre)^2 \second^{-1}}\,,
\end{equation}
which is in good agreement to the expected result according to the Einstein-Stokes equation, Eq. \eqref{eq:diff_ES}. 
The reason for the higher standard deviation of $\unit{0.06}{(\micro\metre)^2 \second^{-1}}$ might be that the fluctuations among the particle radii are larger than indicated by the manufacturer. However, the low diffusion coefficient found in the optical trapping experiment is inside the range of fluctuations of diffusion coefficients among different particles.

\subsection{Comparison between model and data} \label{sec:comparison}

In this section we compare our estimated models to the experimental data. 
At first we compare the single point PDFs. The PDFs of the experimental trajectories are estimated via a standard kernel method using the Epanechnikov kernel \eqref{eq:epanechnikov} and a bandwidth according to Eq. \eqref{eq:silverman}. The PDFs according to the OU model are given by
\begin{equation}
  f(x) = \sqrt{\frac{\gamma}{2\pi \alpha}} \exp\l(-\frac{\gamma}{2\alpha}x^2\r)\,. \label{eq:PDF_OU}
\end{equation}

\begin{figure}[htb]
  \includegraphics[width=0.47\textwidth]{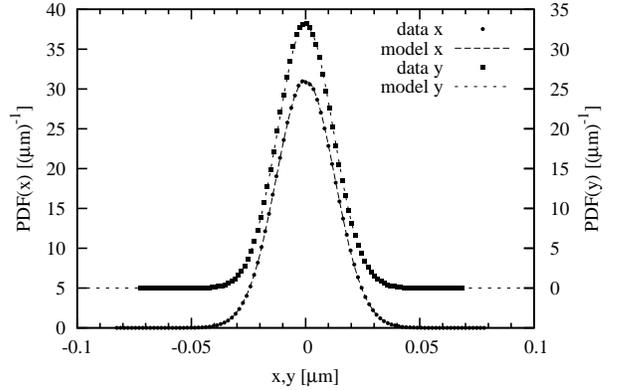}
  \caption{Comparison between the PDFs of the experimental data set (kernel density estimate) and our model (Eq. \eqref{eq:PDF_OU}), for the $x$ component (points and solid line, left vertical axis) and $y$ component (squares and dashed line, right vertical axis) of the particle motion.} \label{fig:pdf_comp}
\end{figure}

Fig. \ref{fig:pdf_comp} shows the estimated PDFs from the experimental data in comparison to the ones from the model time series for the two processes. In both cases the PDFs agree very well.

\begin{figure}[htb]
  \includegraphics[width=0.47\textwidth]{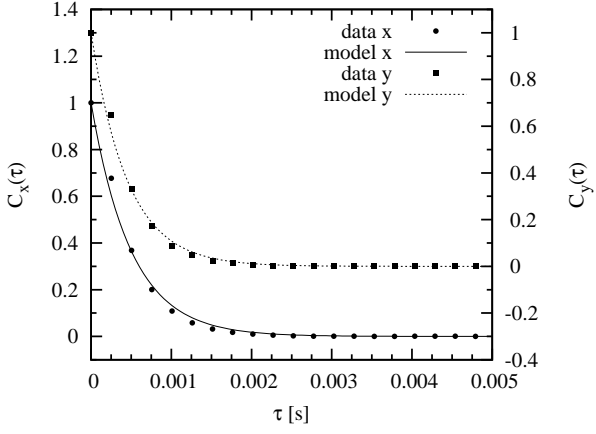}
  \caption{Comparison between the measured autocorrelation functions of the data set and the ACF of the OU model, Eq. \eqref{eq:ACF_OU} for the $x$ component (points and solid line, left vertical axis) and $y$ component (squares and dashed line, right vertical axis) of the particle motion.} \label{fig:ac_comp}
\end{figure}

The second quantity we compare is the ACF, Eqs. \eqref{eq:ACF}. 
Fig. \ref{fig:ac_comp} shows the ACF of the experimental time series for the $x$ and $y$ components of the particle motion, together with ACFs of our model that are given by 
\begin{equation}
    C(\tau) = \exp(-\gamma\tau) \,. \label{eq:ACF_OU}
\end{equation}
As one can also see in Fig. \ref{fig:ac}, the experimental ACFs are not exactly exponentially shaped. The reason for this is discussed in Sec. \ref{sec:conclusion}. However, one can see that the relaxation times of the models and the experimental data sets coincide very well.

\begin{figure}[htb]
  \includegraphics[width=.47\textwidth]{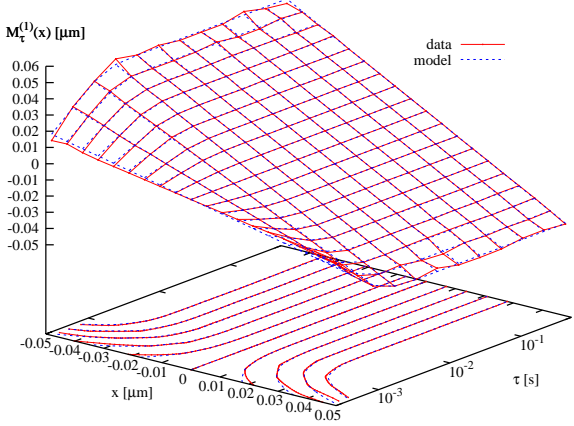}
  \caption{(Color online) Comparison between the first finite time conditional moment of the experimental data set (red solid lines) and the model reconstruction (blue dashed lines) for the $x$ component of the particle motion.} \label{fig:m1_comp}
\end{figure}

\begin{figure}[htb]
  \includegraphics[width=.47\textwidth]{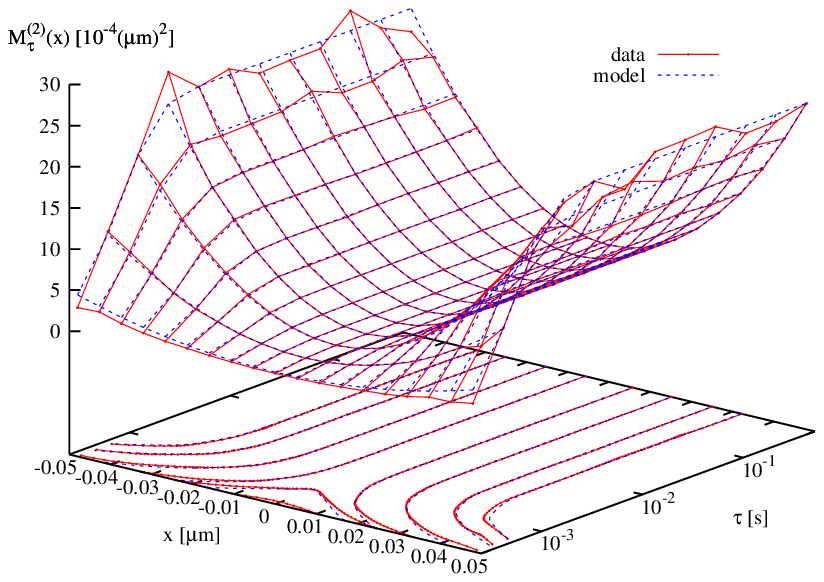}
  \caption{(Color online) Comparison between the second finite time conditional moment of the experimental data set (red solid lines) and the model reconstruction (blue dashed lines) for the $x$ component of the particle motion.} \label{fig:m2_comp}
\end{figure}

\begin{figure}[htb]
  \includegraphics[width=.47\textwidth]{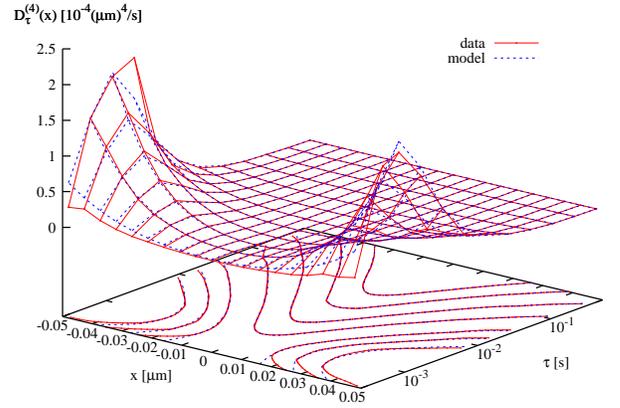}
  \caption{(Color online) Comparison between the fourth finite time KM coefficient of the data set (red solid lines) and the model reconstruction (blue dashed lines) for the $x$ component of the particle motion.} \label{fig:d4_comp}
\end{figure}

As a next step, we compare the first and second finite time conditional moments. They are depicted in Figs. \ref{fig:m1_comp} and \ref{fig:m2_comp}, respectively for the $x$ component. The corresponding figures for the $y$ component are not shown, but are qualitatively equal. The conditional moments of our model are given by
\begin{subequations}
\begin{align}
  M^{(1)}_{\tau}(x) &= -x \l(1-\me^{-\gamma \tau}\r)\,, \\
  M^{(2)}_{\tau}(x) &= x^2\l(1-\me^{-\gamma \tau} \r)^2 + \frac \alpha{\gamma} \l( 1-\me^{-2\gamma \tau} \r)\,.
\end{align}
\end{subequations}
Apparently, our model fits very well to the data. Significant deviations do only occur for the smallest $\tau=\tau_s$, which is below the ME time scale and was not included in the optimization. Here the slope (with respect to $x$) of the first moment and the second moment (for all $x$) of the data are smaller than in the model. An inclusion of time increments below the ME time scale would therefore lead to underestimated drift and diffusion coefficients.

Instead of the fourth conditional moments, we compare the fourth finite time KM coefficients which reads for our model
\begin{align}
  \KM4_{\tau}(x) = &\frac{1}{24\tau} \Bigg[\l(1 - z_{\tau} \r)^4x^4  \nonumber \\
  &+ 6\frac{\alpha}{\gamma}\l(1 - 2z_{\tau} + 2z_{\tau}^3 - z_{\tau}^4 \r)x^2 \nonumber \\
  &+ 3\l(\frac{\alpha}{\gamma}\r)^2\l(1 - z_{\tau}^2 \r)^2 \Bigg] \,. \label{eq:d4}
\end{align}
Here we have used the abbreviation $z_{\tau} = \me^{-\gamma \tau}$. Fig. \ref{fig:d4_comp} shows the estimated fourth KM coefficient together with Eq. \eqref{eq:d4}. As in the two previous plots, significant deviations are only visible for the smallest $\tau=\tau_s$, where the coefficient estimated from data is smaller than in the model. However, 
one can see that $\KM 4_{\tau}(x)$ vanishes as $\tau$ approaches zero. This is the requirement for the Pawula theorem that guarantees that also the third and all higher KM coefficients vanish \cite{risken}.

\section{Conclusion} \label{sec:conclusion}

In the present article we have demonstrated that finite time effects can lead to significant quantitative and also qualitative errors in the KM analysis. A previously published method \cite{honisch2011pre} allows in principle to correct for errors caused by finite time effects. But even the application of this method bears the danger of misinterpretation of the achieved results if the sampling interval of an experimental time series data set approaches the limit of statistical independence discussed in  \cite{anteneodo2010pre}. In order to avoid these misinterpretations, we have extended this method by the calculation of error estimates for the determined model parameters. These error estimates allow for a more honest assessment of the validity of the obtained results. Since the estimated errors only take into account errors caused by finite time effects and the finite amount of available data, they should be regarded as lower bounds for the true model errors, which can also be influenced by other effects.

As long as analytic solutions of the AFPE for the selected parametrization of the KM coefficients are known, the computational effort is very low. The applications presented in this articles have a computation time of less than five seconds on a usual desktop computer.
If the AFPE has to be solved numerically, the computational effort increases dramatically. The MCEP method has not been tested for those cases.

To apply our method to real-world stochastic data, we have also conducted an experiment where trajectories of Brownian particles trapped by an optical tweezers system were measured. We find a ME time scale which is approximately equal to the relaxation time of the process. This can be explained by hydrodynamic memory effects that are still present on this time scale. The large ME time has the consequence that even if the trajectories of the particle were measured with a higher sampling frequency, finite time effects could not be reduced, because time increments below the ME time scale must not be regarded in the KM analysis. 

On time scales above the ME time scale, the process can almost perfectly be reconstructed by an OU process according to the classical overdamped Markov model of Brownian motion. The data quality does not allow to detect deviations from linearity in the drift term or a spatial dependence of the diffusion one could expect because the laser heats up the fluid.

The size of the measured diffusion coefficient is about 1.3 times smaller than the diffusion predicted by the Einstein-Stokes equation.
For comparison, we also measured trajectories of different freely diffusing particles with our experimental setup. Averaged over all particles, the Einstein-Stokes equation was found to be valid. The fluctuations of diffusion constants among different particles, which can probably be traced back to fluctuations among particle radii, are large enough to explain the low diffusion found for the trapped particle.

\begin{acknowledgments}
The authors gratefully acknowledge financial support by Deutsche Forschungsgemeinschaft in the frame of the German-Chinese transregional research cluster TRR 61. C. H. would also like to thank Michael Wilczek and Oliver Kamps for support in connection with the detrending of the data set and Anton Daitche for discussions about the interaction between particles and the surrounding fluid.
\end{acknowledgments}

\bibliography{Lit}

\end{document}